\documentstyle[amssymb,aps,prb,twocolumn,psfig]{revtex}

\draft

\begin{document}

\twocolumn[\hsize\textwidth\columnwidth\hsize\csname@twocolumnfalse\endcsname

\title{Double exchange model for RuSr$_{2}$(Eu,Gd)Cu$_{2}$O$_{8}$}
\author{H. Aliaga and A. A. Aligia}
\address{Comisi\'{o}n Nacional de Energ{\'{\i }}a At\'{o}mica,\\
Centro At\'{o}mico Bariloche and Instituto Balseiro, 8400 S.C. de Bariloche,%
\\
Argentina}
\maketitle

\begin{abstract}
We propose a double exchange model to describe the RuO$_{2}$ planes of RuSr$%
_{2}$(Eu,Gd)Cu$_{2}$O$_{8}$. The Ru$^{+5}$ ions are described by localized
spins, and additional electrons provided by the superconducting CuO$_{2}$
planes are coupled ferromagnetically to them by Hund rules coupling. We
calculate the spin structure factor, magnetic susceptibility and
magnetization as a function of magnetic field and temperature, using a Monte
Carlo algorithm in which the Ru$^{+5}$ spins are treated as classical.
Several experiments which seemed in contradiction with one another are
explained by the theory.
\end{abstract}

\pacs{PACS numbers: 75.30.Vn, 75.10.-b, 75.40.Mg }
]

RuSr$_{2}$(Eu,Gd)Cu$_{2}$O$_{8}$ is a very interesting material because of
the coexistence of magnetic order and superconductivity below $T_{S}\sim 45$
K.\cite{ber,fai} The system orders magnetically at $T_{M}=133$ K, but the
type of magnetic order remains controversial. The first measurements in Gd
samples indicated ferromagnetic (FM) ordering of the Ru moments,\cite{ber}
but neutron diffraction experiments found superlattice reflections
consistent with an antiferromagnetic (AF) order with nearest-neighbor spins
antiparallel in all three directions.\cite{lynn} The Ru contribution to the
magnetic susceptibility at temperature $T>T_{M}$ can be very well described
by $\chi =C/(T-\Theta )$, with $\Theta =100\pm 3$ K.\cite{but} The fact that 
$\Theta >0$ seems difficult to reconcile with AF order at low temperatures.

From previous intensive research on the similar system YBa$_{2}$Cu$_{3}$O$%
_{6+x}$,\cite{ag} it seems clear that each of the superconducting CuO$_{2}$
planes has doping slightly less than 0.1 holes per Cu ion in order to lead
to the observed superconducting critical temperature $T_{S}$. Since Eu or Gd
are expected to be in the oxidation state +3, assuming as usual that the
inactive SrO layers are neutral, charge balance imposes that the RuO$_{2}$
planes have an electron doping $n\lesssim 0.2$ per Ru ion.\cite{ag}
Neglecting covalency with O atoms, this means that there are $1-n$ Ru$^{+5}$
and $n$ Ru$^{+4}$ ions per unit cell. The superconducting CuO$_{2}$ planes
and the magnetic RuO$_{2}$ planes can be regarded as separate entities as a
first approximation. This is supported by symmetry considerations and
band-structure calculations.\cite{pic} We propose that the electronic
structure of the RuO$_{2}$ planes can be described by an effective double
exchange model, with one localized spin at each site representing the Ru$%
^{+5}$ ions, and additional $n$ itinerant electrons per Ru coupled
ferromagnetically with the localized spins.

The model is the two-dimensional (2D) version of one widely used in
manganites:

\begin{equation}
H=-\sum_{\langle ij\rangle \sigma }t\left( c_{i\sigma }^{\dag }c_{j\sigma }+%
{\rm H.c.}\right) -J_{H}\sum_{i}{\bf s}_{i}{\bf .S}_{i}+K\sum_{\langle
ij\rangle }{\bf S}_{i}{\bf .S}_{j},  \label{h}
\end{equation}
Here ${\bf S}_{i}$ is the localized spin operator representing the Ru$^{+5}$
ion at site $i$, $c_{i\sigma }^{\dag }$ is the operator creating an
itinerant electron of spin $\sigma $ at this site, and ${\bf s}%
_{i}=\sum_{\alpha \beta }c_{i\alpha }^{\dag }\sigma _{\alpha \beta
}c_{i\beta }$ gives the spin of this electron. For simplicity we take the
Hund coupling, $J_{H}\rightarrow +\infty $.

The model is solved using a classical Monte Carlo (MC)\ procedure for the
localized spins, in conjunction with exact diagonalization of the conduction
electron system.\cite{yuno} The localized spins are taken to be classical
and of magnitude one. The conduction electrons are assumed to occupy a
single orbital, and from the condition $J_{H}\rightarrow +\infty $ only one
spin projection is possible at each site. Then, for each configuration of
localized spins, the effective hopping $t_{ij}$ of itinerant electrons
between two sites $i$ and $j$ is affected by the factor $\langle \uparrow
_{i}|\uparrow _{j}\rangle $, where $|\uparrow _{j}\rangle $ is the state of
a spin 1/2 pointing in the direction of the localized spin $j$. Then: 
\begin{equation}
t_{ij}=t\left( \cos \frac{\theta _{i}}{2}\cos \frac{\theta _{j}}{2}%
+e^{-i(\phi _{i}-\phi _{j})}\sin \frac{\theta _{i}}{2}\sin \frac{\theta _{j}}{%
2}\right) ,  \label{edet}
\end{equation}
where $\theta _{i}$ and $\phi _{i}$ are the polar angles of spin ${\bf S}_{i}
$. The resulting electronic energy levels are then filled by the available
number of electrons in the canonical ensemble. The MC simulation proceeds
from the partition function with classical spins. The general features of
this model as a function of $K/t$ and doping have been studied before in the
context of the manganites.\cite{hor,sev} The calculations reported here are
in a square of $N=$64 atoms with periodic boundary conditions and six mobile
electrons ($n=6/64\cong 0.1$). We have taken $K/t=0.06$. We expect that the
order of magnitude of $t$ is near 0.25 eV, and then $K\sim 180$ K, near the
observed $T_{M}$.

We have calculated the spin structure factor defined as $S({\bf q}%
)=\sum_{i,j}{\bf <}{\bf S}_{i}{\bf .S}_{j}{\bf >}e^{i{\bf q.}({\bf r}_{i}-{\bf r}%
_{j})}/N^{2}$ for all nonequivalent values of ${\bf q}$ in our $8\times 8$
cell. In Fig. 1 we represent the two superlattice peaks of highest intensity
as a function of temperature. Clearly, the peak at ${\bf q}=(\pi ,\pi )$ is
the dominant one, in agreement with neutron experiments.\cite{lynn} An
important fact is that at $T=0$ the amplitude of $S({\bf q})$ is smaller
than one. This is consistent with experiments because the magnetic moment
deduced from magnetic measurements \cite{but} is larger than that deduced
from the amplitude of the superlattice reflections. \cite{lynn} This
reduction in our simulations is due to the fact that in general, there are
small ferromagnetic islands (magnetic polarons) around the mobile electrons.
Since these islands are not entirely compensated in our small systems and
that their kinetics is slow at low temperatures, they give rise to a small
contribution to $S(0,0)$ at small temperatures, as displayed in Fig. 1.
Since we are dealing with a strictly 2D system in our simulations (instead
of 3D in the real system), and a continuous symmetry cannot be broken in 2D,
Fig. 1 does not show a phase transition to the paramagnetic phase.

\begin{figure}[t]
\centerline{\psfig{figure=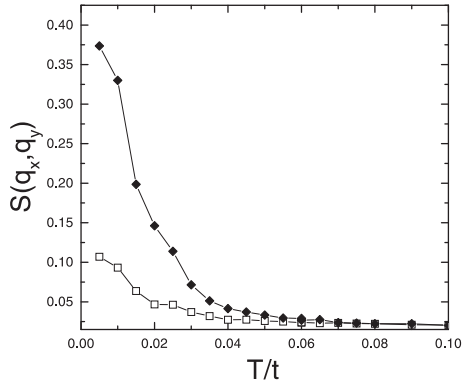,height=5cm,angle=0}}
\medskip
\caption{Spin structure factor as a function of temperature for two wave 
vectors ${\bf q}$: ${\bf q}=(\pi ,\pi )$ (diamonds), ${\bf q}=(0,0)$ 
(squares).}
\end{figure}

The mobility of the ferromagnetic polarons makes it easy to orient them
under small applied magnetic fields. Then, as observed experimentally \cite
{but} the magnetization increases abruptly with small magnetic fields ($\sim
5$ T) compared with the Neel temperature $T_{M}\cong 133$ K. This fact is
unusual for an ordinary AF system. The magnetization curve of the localized
spins in our system is shown in Fig. 2. In spite of the difficulties with
the sluggish statistics of the ferromagnetic polarons mentioned above, one
can see an abrupt increase at low fields, due to the alignment of these
polarons, followed by a linear increase with a smaller slope, in agreement
with experiment.\cite{but}.

Another unexpected result for an AF system is the fact that the inverse
magnetic susceptibility $\chi ^{-1}$ at high temperatures extrapolates to a
positive temperature for $\chi ^{-1}=0$. As shown in Fig. 3, our model also
reproduces this result. The constant $\Theta $ obtained fitting $\chi
=C/(T-\Theta )$ is $\Theta =(0.041\pm 0.004)t$. This is a subtle issue which
is related with the way in which the ferromagnetic polarons are reduced as
the temperature is increased, and their effect on the magnetic correlations
on their neighborhood. A mean field picture in which the kinetic energy of
the carriers is neglected seems unable to explain this result even
qualitatively. The resulting value of the slope gives $C=(0.314\pm 0.006)\mu
^{2}$, where $\mu $ is the magnetic moment of the localized spins. This
value is near to the classical value $\mu ^{2}/3$ for noninteracting spins.
The value of $C$ in the real system depends on the spin of the
configurations 4d$^{5}$ and 4d$^{6}$ of Ru$^{+5}$ and Ru$^{+4}$.\cite{but}

\begin{figure}[t]
\centerline{\psfig{figure=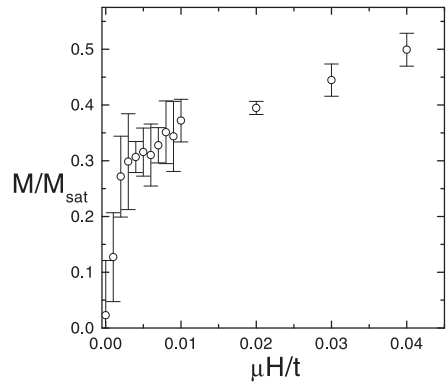,height=5cm,angle=0}}
\medskip
\caption{Magnetization as a function of applied magnetic field at
temperature $T=0.02K$. $M_{sat}$ is the saturation magnetization.}
\end{figure}

\begin{figure}[t]
\centerline{\psfig{figure=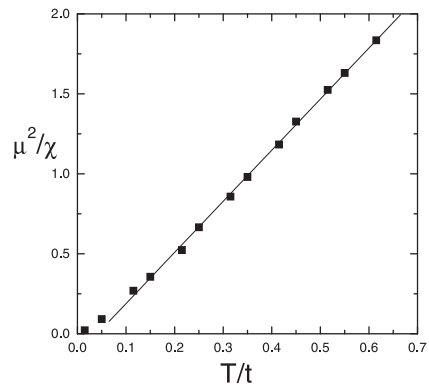,height=5cm,angle=0}}
\medskip
\caption{Inverse magnetic susceptibility as a function of temperature.
The straight line is a fit using $\chi ^{-1}=(T-\Theta )/C$.}
\end{figure}

To conclude, we have proposed a double exchange model for RuSr$_{2}$(Eu,Gd)Cu%
$_{2}$O$_{8}$ which is able to explain the main unusual features of these
systems. Previous studies of the model \cite{hor,sev} indicate that for
small densities there is phase separation in the model. However, in the real
system, particularly due to the small number of carriers \cite{ag}, the
effect of long-range Coulomb repulsions, not included in the model should be
present and avoid macroscopic phase separation,\cite{coul} while phase
separation in a microscopic scale does not affect our conclusions.

We thank B. Alascio for useful discussions. We are partially supported by
CONICET. This work was sponsored by PICT 03-06343 of ANPCyT and PIP 4952/96
of CONICET.

\end{document}